\documentstyle[12pt,psfig]{iopart}
\begin{document}

\jl{1}

\title{Stretched exponentials and power laws in granular avalanching}

\author{D A Head\dag
\footnote[2]{Present address: Department of Physics and Astronomy,
JCMB King's Buildings, University of Edinburgh, Edinburgh EH9 3JZ,
United Kingdom. E-mail: david@ph.ed.ac.uk}
and G J Rodgers\S
\footnote[4]{E-mail: G.J.Rodgers@brunel.ac.uk}}

\address{\dag Institute of Physical and Environmental Sciences,
Brunel University, Uxbridge, Middlesex, UB8~3PH, United Kingdom}

\address{\S Department of Mathematics and Statistics,Brunel University,
Uxbridge, Middlesex, UB8~3PH, United Kingdom}

\begin{abstract}
We introduce a model for granular surface flow
which exhibits both stretched exponential and power law
avalanching over its parameter range.
Two modes of transport are incorporated,
a rolling layer consisting of
individual particles and the overdamped, sliding
motion of particle clusters.
The crossover in behaviour observed
in experiments on piles of rice is attributed to
a change in the dominant mode of transport.
We predict that power law avalanching
will be observed whenever surface flow is
dominated by clustered motion.
\end{abstract}


\maketitle


Although the sandpile model was introduced some time ago~\cite{BTW},
its relevance to real granular materials remains unclear.
It predicts that a pile formed by adding particles
one at a time to a flat surface with open boundaries
will naturally evolve to a
continuous phase transition,
illustrating the concept of {\em self--organised criticality}\/ or SOC.
The signal that such a critical state has been reached
is when the spectrum of avalanche sizes becomes scale-invariant;
that is, {\em power law}\/ in form.
By contrast, experiments using sand and beads
have shown that, if anything,
the generic form for the
avalanche spectrum is {\em stretched exponential}\/
rather than power law~\cite{exp1}---\cite{feder}.
Whatever else the sandpile model may be, it is clearly
not~a very good model for piles of sand.

A cursory observation of real sandpiles in action
easily reveals the shortcomings of the model.
Inertia has been neglected~\cite{inertia1}---\cite{inertia3}
and the particles are assumed to move purely by a series of
toppling events, defined in terms of the relative instabilities
of local regions of the pile's surface.
In fact, real particles
gain kinetic energy as they accelerate downslope,
dislodging other particles from the static bulk
which combine to form a {\em rolling layer}.
An intuitively more plausible approach treats this
rolling layer as being governed by a convective diffusion
equation~\cite{BCEP}, subsequently likened to a fluid layer
interacting with the solid bulk~\cite{rev}.
This seems to be more in line with the experiments, for
instance in predicting hysteresis in the variation of the
angle of repose.

The first round of experiments all involved roughly spherical particles
that could easily roll under their own weight once activated.
Recently, similar experiments have been performed using highly
anisotropic particles instead, namely grains of rice~\cite{nature}.
For rounder grains the avalanche spectrum
was again found to be stretched exponential, but for
grains with a large aspect ratio a very different picture emerged.
These grains did not often roll but
tended to slide along the surface in coherent domains,
hereafter referred to as {\em clusters}~\cite{mehta}---\cite{cluster3}.
The corresponding avalanche spectrum was power law.
This rekindled hopes that some granular systems might after
all be SOC, and a number of modified sandpile models
have now been devised~\cite{ricepile1}---\cite{oslo2}.
However, these ``ricepile'' models lack a clear physical interpretation
and fail to exhibit stretched exponential behaviour
over any part of their parameter space.
Furthermore, none of them have the same exponent for the
avalanche spectrum as those obtained from the experiments.
This has led to speculation about whether the power law
really is a consequence of SOC behaviour~\cite{cohnoise}.

In this paper, we present a picture for granular avalanching
which closely follows the qualitative descriptions of surface
transport made during the ricepile experiments~\cite{nature}.
A simple model is devised which describes the transport process
on the level of clusters.
Numerical simulations show that the avalanche spectrum is stretched exponential
for a broad parameter range, but crosses over to a power law
regime when the surface transport is principally in the form of clusters.
The recovery of SOC behaviour is hence attributed to the
overdamped motion of clusters as opposed to the inertial
motion of individual grains.

\addtocounter{section}{1}
\section{A simple model for the dual transport mechanism}

We now proceed to derive a model which incorprates both
individual particle and clustered motion.
Although clearly very simplified, this may be regarded as
the minimal model for dual surface flow.

Only the clusters themselves will be explicitly represented;
independent grains are assumed to form a rolling layer
that will be implicitly incorporated into the rules of the model.
To approximate the discrete nature of clusters,
a lattice repesentation is adopted in which the
width of the pile is given by the integer variable~$i$,
\mbox{$1\leq i\leq L$}.
The boundary at \mbox{$i=1$} is closed and clusters can only exit
the system via the open boundary at~\mbox{$i=L$}.
On each site $i$ are stacked $N_{i}$ clusters of
varying height ${\rm dh}_{ij}$\,, \mbox{$1\leq j\leq N_{i}$}\,,
so the total height is

\begin{equation}
h_{i} = \sum_{j=1}^{N_{i}}{\rm dh}_{ij}
\end{equation}

\noindent{}and the local slope is \mbox{$z_{i}=h_{i}-h_{i+1}$}
(it is customary in sandpile
models to use positive slopes even though the height is
decreasing in the direction of increasing $i$).
This discrete representation is similar to that adopted by
other sandpile models, except that here the blocks represent
whole clusters rather than just single particles.

\begin{figure}
\centerline{\psfig{file=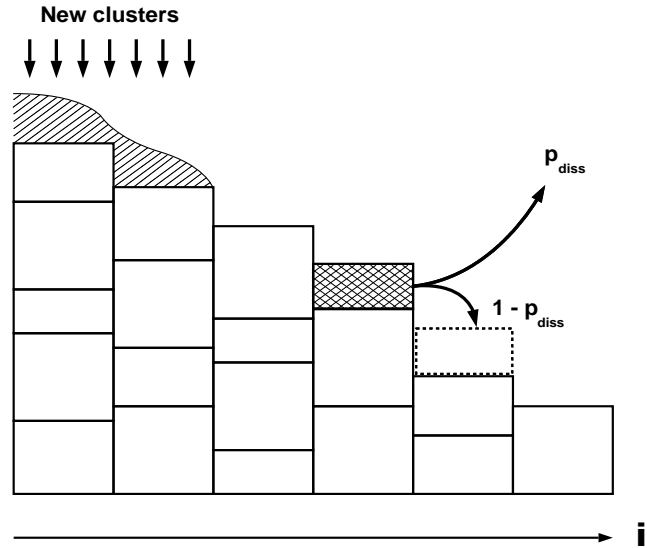,width=3.5in}}
\caption{A schematic representation of the
influx and movement of clusters.
New clusters can only form on the surface in regions
where the rolling layer (here denoted by the
diagonally shaded area) is non--vanishing.
An existing cluster begins sliding once the local
slope exceeds some given threshold value.
For instance, if the cross--hatched cluster in the diagram
becomes activated, it
either dissipates into the rolling layer with
probability $p_{\rm diss}$ or moves one site to the
right with probability \mbox{$1-p_{\rm diss}$}.
}
\label{f:rlayer}
\end{figure}

New clusters can emerge in regions over the surface where the rate
of conversion from rolling to static particles is non-zero.
Anisotropic particles quickly lose their kinetic energy via
uneven rolling {\em etc.} and consequently the rolling layer
is limited to the vicinity of the closed boundary.
In this case new clusters only arise near to \mbox{$i=1$},
as schematised in Fig.~\ref{f:rlayer}.
By contrast particles that easily roll, such as the rounder grains
of rice used in the experiments, gives rise to a rolling layer extends
deep into the system, so clusters may emerge anywhere over the pile's surface.
Note that clusters may emerge on sites \mbox{$i>1$} even though the
individual particles are only added next to the closed boundary.

Clusters on the surface may move under perturbations from
the rolling layer or the motion of adjacent clusters.
To model this, a cluster at site $i$ whose local slope $z_{i}$
exceeds some threshold value $(z_{c})_{i}$ becomes active and
begins to move.
Following Christensen {\em et al.}\/, $(z_{c})_{i}$ is taken to be
a site-dependent random variable that changes value after
each sliding event~\cite{transit}---\cite{oslo2}.
This annealed disorder represents the disordered packing of
granular media.
In the model specified below we also incorporate a number
of other forms of disorder, such as the size of clusters
and the number of clusters that move at once.
Although the critical exponents vary for small systems,
we were unable to simulate large systems and so it is
possible that all of these models belong to the
same universality class of one dimensional sandpile models with
annealed disorder~\cite{ricepile1}---\cite{oslo2}.

A moving cluster is large compared to the individual particles that
activated it, so its velocity will be low and hence its motion will
be {\em overdamped}.
This means it will only move one site before coming to rest.
However, in our model there is a fixed probability $p_{\rm diss}$ per sliding
event that the cluster disintegrates into its constituent
particles and disperses into the rolling layer,
which on this level of description is equivalent to dissipation.
For incompressible particles \mbox{$p_{\rm diss}=1$},
which may relate to the {\em fragility}\/ of
structures formed by such materials~\cite{fragile}.
Conversely, \mbox{$p_{\rm diss}<1$} for deformable materials,
decreasing still further
for elongated or jagged grains as a consequence of
their increased contact surface area.
Note that the dissipation of clusters into the rolling layer
does {\em not}\/ imply loss of mass conservation.

In summary, the model is specified as follows.
Each site $i$ is assigned a critical slope variable $(z_{c})_{i}$\,.
For every time step, the following procedure is performed.

\vspace{0.2in}

\noindent{}(i) {\em Driving:}
A site $i$ is selected and a cluster of height ${\rm dh}_{ij}$
is added to the top of it.

\vspace{0.1in}

\noindent{}(ii) {\em Check for stability:} Any site $k$ whose local
slope \mbox{$z_{k}>(z_{c})_{k}$} is marked for toppling.

\vspace{0.1in}

\noindent{}(iii) {\em Toppling:}
All of the unstable sites marked in step (ii) are toppled in parallel.
For each~$k$, the topmost $n$ clusters are selected and moved to
site $k+1$, or are removed from the system with probability $p_{\rm diss}$\,.

\vspace{0.1in}

\noindent{}(iv) {\em Annealed $z_{c}$:}
Every site that toppled is assigned a new critical slope $z_{c}$\,.

\vspace{0.1in}

\noindent{}(iv) {\em Avalanche:}
Steps (ii)--(iv) are repeated until every site in the system is
stable, then another cluster is added as per step~(i).

\vspace{0.2in}

\noindent{}The distributions for the random variables
${\rm dh}_{ij}$\,, $z_{c}$, $n$ and the site $i$ where the clusters
are added need to be specified
in some manner that relates to the type of grain involved. This will
be described in the following section.

\section{Comparison with experimental data}

It will now be demonstrated that this model exhibits
the same crossover in behaviour as in the ricepile
experiments.
The parameter $p_{\rm diss}$ and the range of cluster addition
depend upon the properties of the type of grain in question,
in a manner to be described below.
The other parameters used in the simulations were the same for both
cases.
The cluster heights ${\rm dh}_{ij}$
were drawn from the uniform distribution $[a,b]$.
The threshold slopes $z_{c}$ were uniformly distributed in $[2,3]$,
although other ranges were also considered with
no qualitative change in behaviour observed.
When a site becomes unstable the topmost $n$ clusters
are activated and start to move,
where $n$ is either fixed at 1 or randomly selected
from~\mbox{$\{1,2\}$}.
No significant changes in behaviour were observed
for all reasonable parameter ranges,
although deviations did occur in some
extreme cases, for instance when~\mbox{$b\gg a$}
or~\mbox{$a\rightarrow0$}.

To simulate a system comprising of the rounder grains
that gave rise to a broad rolling layer but
did not form stable clusters,
new clusters are added to the surface uniformly
over the range \mbox{$1\leq i\leq L$}, and
\mbox{$p_{\rm diss}\in(0.1,0.6)$}.
A new cluster is not added until all of the sites
in the system are stable.
Numerical simulations measuring $P(E,L)$, the distribution of
the potential energy $E$ lost by the pile as the result of
a single addition,
shows good data collapse after the finite--size rescaling
\mbox{$P(E,L)=L^{-1}f(E/L)$},
in accord with the experimental data.
The scaling function $f(x)$ is a stretched exponential,

\begin{equation}
f(x)=A\exp\{-(x/x_{0})^{\gamma}\}
\end{equation}

\begin{figure}
\centerline{\psfig{file=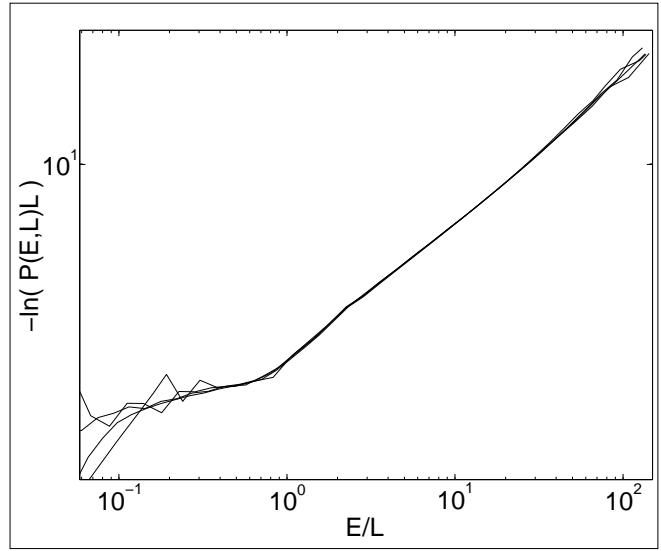,width=4in}}
\caption{Finite--size scaling plot of $P(E,L)$ for
system sizes \mbox{$L=100$}, 250, 500 and 1000.
New clusters were added
uniformly over the entire system and~\mbox{$p_{\rm diss}=0.4$}.
If \mbox{$f(x)=A\exp\{-(x/x_{0})^{\gamma}\}$}
then $-\ln f(x)=-\ln A+(x/x_{0})^{\gamma}$,
which will give a straight
line on a log--log plot when $x$ is large and
the corrections due to $\ln A$ can be ignored.
For the plots shown above, $\gamma\approx0.43$ and $x_{0}\approx0.4$,
in agreement with the values from the ricepile experiments,
\mbox{$\gamma=0.43\pm0.03$} and \mbox{$x_{0}=0.45\pm0.09$}.
The other parameters were \mbox{$z_{c}\in[2,3]$}, \mbox{$n\in\{1,2\}$},
and \mbox{${\rm dh}_{ij}\in[0.5,1.5]$}.
Units were chosen such that~\mbox{$mg=1$}.
}
\label{f:strexp}
\end{figure}

\noindent{}with~\mbox{$0<\gamma<1$},
where $\gamma$ and $x_{0}$ depend upon the parameters.
By varying $p_{\rm diss}$ we found it easy to obtain
values close to those measured in the early experiments
and the ricepile experiments~\cite{feder,nature}, as illustrated
in Fig.~\ref{f:strexp}.

\begin{figure}
\centerline{\psfig{file=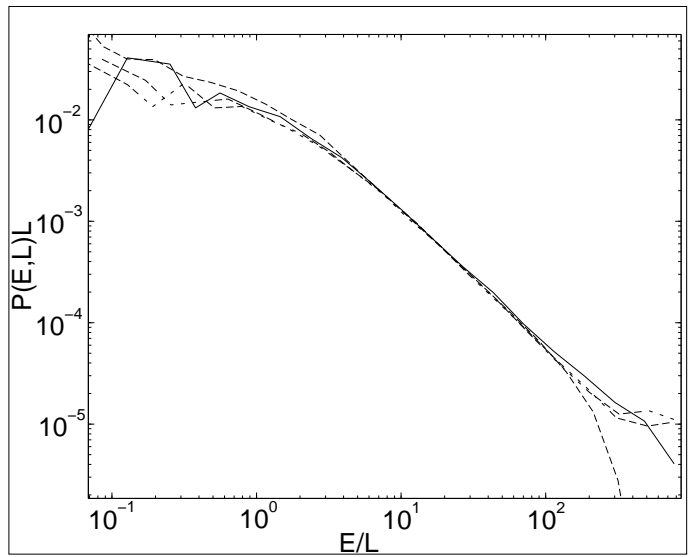,width=4in}}
\caption{Finite--size scaling plot of $P(E,L)$ for four different
system sizes,
\mbox{$L=500$} (dotted line), \mbox{$L=1000$} (dashed-dotted
line), \mbox{$L=1500$} (dashed line) and \mbox{$L=2000$} (solid line).
New clusters were added uniformly over the range
\mbox{$1\leq i\leq L/20$} and \mbox{$p_{\rm diss}=0$}.
The straight line has a slope of approximately \mbox{$1.30\pm0.05$}.
The other parameters were
\mbox{$z_{c}\in[2,3]$}, \mbox{$n\in\{1,2\}$}
and \mbox{${\rm dh}_{ij}\in[0.8,1.2]$}.
}
\label{f:clust}
\end{figure}

The elongated grains of rice formed stable clusters that
passed intact through the system, but the penetration
of the rolling layer was limited.
This translates into
\mbox{$p_{\rm diss}=0$} and new clusters
only being added to sites $i$ in a range such as
\mbox{$1\leq i\leq L/10$} or \mbox{$1\leq i\leq L/20$}.
$P(E,L)$ obeys the same finite size scaling relation as before,
but $f(x)$ is now flatter for small $x$ and power law
for large $x$, \mbox{$f(x)\sim x^{-\alpha}$}, as
in Fig.~\ref{f:clust}.
The exponent $\alpha$ depends upon the choice of parameters and the
system size, typically taking values in the range 1.1 to 1.6.
We were unable to obtain an exponent close to the experimental
value, which was ``just greater than 2''~\cite{nature}.
This might be due to some crucial dynamical effect which
has been overlooked, although it may just be an
artefact of the reduced dimensionality of the model.
Since \mbox{$\alpha<2$} a finite--size cut--off
is necessary and large systems
are needed for a definite power law region to appear.
However, measurements of the
number of clusters that moved during each avalanche,
rather than the change in potential energy,
shows a clearer power law
(with the same exponent $\alpha$) for all system sizes.
Coupled with the similarity to many sandpile models, this
implies that the model is SOC in this region of parameter space.

\begin{figure}
\centerline{\psfig{file=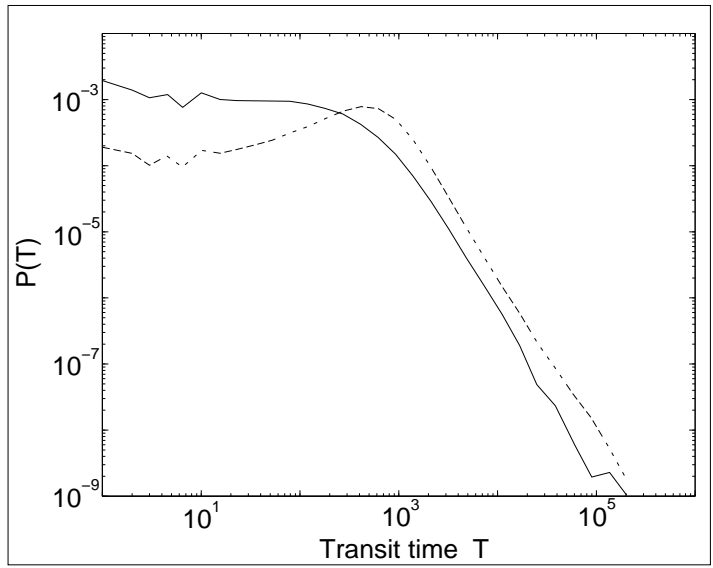,width=4in}}
\caption{The distribution of transit times for clusters
to move through the system, for the same parameter
values as those given in
Fig.~2 (solid line) and Fig.~3
(dashed line).
The straight line has a slope of \mbox{$2.3\pm0.1$}.
The system size was \mbox{$L=250$} in both cases.
}
\label{f:transit}
\end{figure}

Christensen {\em et al.} performed
a further set of experiments to measure the time taken
for tracer particles (coloured grains of rice) to pass
through the system~\cite{transit}.
Only the elongated grains that gave
the power law in the first set of experiments were used.
They found that the distribution of transit times
$P(t)$ was roughly constant for small $t$ but crossed over
to a power law for large~$t$, \mbox{$P(t)\sim t^{-\beta}$}
with \mbox{$\beta=2.4\pm0.2$}.
Simulations of our model show the same behaviour
with a similar exponent for {\em both}\/ cases studied,
as demonstrated in Fig.~\ref{f:transit}.
Thus we conclude that this power law is {\em not}\/
symbolic of a critical state and predict that the same
exponent would be recovered if the experiments were repeated
using the rounder grains.

\section{Summary}

In summary, we have argued that the
crossover in behaviour observed
in the ricepile experiments can be attributed to
a change in the dominant mode of transport,
from the inertial motion of the rounder grains
to the overdamped motion of clusters of the elongated grains.
A simple model was introduced in which blocks represent
coherent clusters of particles.
Not only does this explain why inertial effects may be ignored,
since clusters naturally move in an overdamped fashion, but also allows for
the dissipation of blocks in a physically plausible manner.
This may be a important mechanism for the emergence
of the stretched exponential behaviour.

If this hypothesis is correct, then power law avalanching
should also be observed for any material in which surface flow is
dominated by clustered motion.
It would be interesting to see if this prediction
might be verified experimentally.
We have recently become aware of experiments on watered
soil ridges where regions of the surface moved as whole regions~\cite{soil}.
Power law behaviour was observed, which would appear to be
in accord with our hypothesis.



\end{document}